\documentclass[trackchanges]{aastex62}

\usepackage{xspace}
\usepackage{amsmath}

\newcommand{\ang}{\AA\xspace}

\newcommand{\tw}{\ensuremath{\xi}\xspace}
\newcommand{\td}{\ensuremath{f}\xspace} 
\newcommand{\tdmax}{\ensuremath{f_{\mathrm{max}}}\xspace} 
\newcommand{\tdmean}{\ensuremath{f_{\mathrm{mean}}}\xspace}
\newcommand{\gammantwo}{\ensuremath{\Gamma \left[ \frac{n}{2} \right]}}
\newcommand{\nd}{\ensuremath{n}\xspace}
\newcommand{\NSNe}{\ensuremath{N_{\mathrm{SNe}}}\xspace}
\newcommand{\We}{I\xspace}
\newcommand{\we}{I\xspace}

\newcommand{\Our}{My\xspace}

\newcommand{\nsixth}{\ensuremath{4.7^{+4.5}_{-2.1}}\xspace}
\newcommand{\sigsixth}{\ensuremath{0.9^{+0.6}_{-0.4}}\xspace}
\newcommand{\meanysixth}{\ensuremath{0.051^{+0.006}_{-0.006}}\xspace}
\newcommand{\ymaxsixth}{\ensuremath{0.14^{+0.03}_{-0.03}}\xspace}
\newcommand{\ntwelveth}{\ensuremath{3.2^{+1.9}_{-1.1}}\xspace}
\newcommand{\sigtwelveth}{\ensuremath{0.8^{+0.4}_{-0.3}}\xspace}
\newcommand{\meanytwelveth}{\ensuremath{0.103^{+0.010}_{-0.009}}\xspace}
\newcommand{\ymaxtwelveth}{\ensuremath{0.22^{+0.03}_{-0.03}}\xspace}
\newcommand{\neighteenth}{\ensuremath{2.8^{+1.5}_{-1.0}}\xspace}
\newcommand{\sigeighteenth}{\ensuremath{0.9^{+0.4}_{-0.3}}\xspace}
\newcommand{\meanyeighteenth}{\ensuremath{0.156^{+0.013}_{-0.012}}\xspace}
\newcommand{\ymaxeighteenth}{\ensuremath{0.30^{+0.03}_{-0.03}}\xspace}

\newcommand{\roundedn}{5, 3, and 3, respectively\xspace}
\newcommand{\bestestimatesofn}{three to five\xspace}

\graphicspath{{./}{figures/}}

\shorttitle{SN Ia Dimensionality}
\shortauthors{Rubin}

\begin{document}

\title{Constraining the Dimensionality of SN Ia Spectral Variation with Twins}

\correspondingauthor{David Rubin}
\email{drubin@stsci.edu}

\author[0000-0001-5402-4647]{David Rubin}
\affiliation{Space Telescope Science Institute, 3700 San Martin Drive, Baltimore, MD 21218}
\affiliation{E.O. Lawrence Berkeley National Laboratory, 1 Cyclotron Rd., Berkeley, CA, 94720}

\begin{abstract}

SNe Ia continue to play a key role in cosmological measurements. Their interpretation over a range in redshift requires a rest-frame spectral energy distribution model. For practicality, these models are parameterized with a limited number of parameters and are trained using linear or nonlinear dimensionality reduction. This work focuses on the related problem of estimating the number of parameters underlying SN Ia spectral variation (the dimensionality). \We present a technique for using the properties of high-dimensional space and the counting statistics of ``twin'' SNe Ia to estimate this dimensionality. Applying this method to the supernova pairings from \citet{fakhouri15} shows that a modest number of parameters (\bestestimatesofn, not including extinction) explain those data well. The analysis also finds that the intrinsic parameters are approximately Gaussian-distributed. The limited number of parameters hints that improved SED models are possible that may enable substantial reductions in SN cosmological uncertainties with current and near-term datasets.

\end{abstract}

\keywords{supernovae: general, methods: statistical}

\section{Introduction} \label{sec:intro}

Type Ia supernovae (SNe Ia) continue to be developed and applied as ``standardizable candles,'' enabling them to be used as distance measures over cosmological distances. Observations of SNe Ia provided the first strong evidence for an accelerated expansion rate \citep{riess98, perlmutter99}. With larger sample sizes, redshift reach, and better measurement quality, these SNe continue to inform our understanding of the dynamics of the universe \citep{scolnic18, riess18, des18}.

In order to use SNe Ia for cosmological measurements over a range of redshifts, we must fit a rest-frame spectral energy distribution (SED) model to the data for each SN. At this point, there are several SED models for SNe Ia, with different numbers of parameters. This work explores a complementary question related to training these models: how many parameters drive SN Ia SED variation, i.e., what is the dimensionality of the parameter space? Constraining this dimensionality has at least a few applications.

\newcommand{\AppTwo}{{\it To set a scale on the data required for future cosmological applications.}\xspace}
\newcommand{\AppThree}{{\it To assess the performance of current empirical models of SED variation and selection effects.}\xspace}
\newcommand{\AppOne}{{\it To inform an enumeration of the physical parameters.}\xspace}
\newcommand{\AppFour}{{\it To semi-quantitatively inform simulations of SNe Ia by probing the physical-parameter distribution functional form, separate from dimensionality.}\xspace}

\begin{enumerate}
\item \AppThree Establishing that there are more SN parameters than model parameters indicates a deficiency in the SED model training or a major selection effect in the training data. Establishing fewer SN parameters than SED model parameters indicates an inefficient SED parameterization.

\item \AppTwo It is very likely that more parameters requires more detailed data (on at least a training subset of SNe) to enable precision cosmological constraints. Such detailed data might include higher signal-to-noise or better-sampled light curves, more wavelength coverage, or high-quality spectroscopy. As the next generation of SN cosmology experiments (e.g., the {\it Wide Field Infrared Survey Telescope}, \citealt{spergel15}, and the Large Synoptic Survey Telescope, \citealt{lsst09}) is currently in the planning phase, getting an indication of the dimensionality is timely.
\end{enumerate}

More speculatively, the parameters underlying spectral variation may be cleanly related to physical parameters of the progenitor and its explosion. In that case, establishing the dimensionality has other applications.

\begin{enumerate}
\setcounter{enumi}{2}
\item \AppOne 
\item \AppFour 
\end{enumerate}

Frequently, we parameterize SNe Ia with linear dimensionality reduction. A frequently used example is the Spectral Adaptive Light-curve Template (SALT2) \citep{guy07}, which models SN flux\footnote{Here, flux has dimensions of energy per time$\cdot$area$\cdot$wavelength, e.g., ergs s$^{-1}$ cm$^{-2}$ \ang$^{-1}$.} as a function of phase with respect to maximum light ($p$) and rest-frame wavelength ($\lambda$) as the sum of a mean time-evolving SED ($m_0$) and a linear correction ($m_1$):
\begin{equation}
f(p,\ \lambda;\ x_0,\ x_1,\ c) = x_0 \cdot [m_0(p,\ \lambda) + x_1 \cdot m_1(p,\ \lambda)] \cdot e^{CL (\lambda) \cdot c} \;,
\end{equation}
where CL is color law and $c$ is the color. Other parameterized classifications of SNe are based on spectral features, such as the two-parameter pseudo equivalent width (pEW) of Si {\sc ii} $\lambda$5972 \ang and Si {\sc ii} $\lambda$6355 \ang \citep{branch06}. More recently, \citet{saunders18} presented SuperNova Empirical MOdels (\textsc{SNEMO}), which have versions with one linear component of variation (\textsc{SNEMO2}, similar to SALT2), six components (\textsc{SNEMO7}), and fourteen (\textsc{SNEMO15}), but the expansion is still linear in flux. Such linear dimensionality reductions will require extra components to describe nonlinear processes.\footnote{For example, data consisting of a simple shift in centroid of a line profile requires an infinite number of linear principal components to describe, although the data are generated with one parameter that has a nonlinear impact on flux.} Thus they cannot be used to estimate the dimensionality of SNe Ia. However, we might expect that fourteen is a reasonable upper limit to the dimensionality of SNe Ia, at least in the rest-frame optical for SNEMO's phase range of $-10 < p < 46$ with respect to maximum light.

A related class of approaches is nonlinear dimensionality reduction, e.g., locally linear embedding \citep{roweis00} or Isomap \citep{tenenbaum00}. However, these approaches are difficult to apply to SN Ia data. Both approaches cannot handle measurements that are not near any other measurements (peculiar SNe), where there is no well-defined local geometry to learn. Further, Isomap requires interpretation of differences between similar SNe as distances, which presents a further challenge. With increased sample sizes (which may connect isolated SNe to the bulk) and better metrics for comparing SNe, both of these limitations will be significantly addressed in the future \citep{boone19}.

\Our approach here is to probe the dimensionality of the SN Ia parameter space using the counting statistics of ``twin'' SNe Ia, SNe that look similar. Section~\ref{sec:approach} motivates this approach. Section~\ref{sec:inference} describes the details of the inference while Section~\ref{sec:simulated} presents an evaluation of that inference on simulated data. Section~\ref{sec:dicussion} presents the dimensionality constraints inferred from the \citet{fakhouri15} data, and performs alternative related analyses. Section~\ref{sec:conclusions} discusses the implications for the questions raised in this section, considers next steps, and concludes. Finally, Appendix~\ref{sec:derivation} presents the derivation of the analytic PDF.

\section{Analysis}

\subsection{The Twins-Based Approach} \label{sec:approach}

\citet{fakhouri15} took a nonlinear but unparameterized approach to modeling SNe: matching twins. They took optical spectrophotometric time series collected by the Nearby Supernova Factory \citep{aldering02}, interpolated each SN in a pair to the other SN's phases, and compared them (over all phases in one set of analyses, and at maximum light in others). SNe pairs that matched better than a given threshold in ``twinness'' (\tw, a pseudo-$\chi^2$ that also takes differential extinction into account) were identified as twins, and in fact showed a much smaller dispersion in luminosities than SN pairs with large mutual \tw values. \citet{fakhouri15} placed the twins in 17 bins of \tw (each one containing 6\% of the SN pairs), and this work considers the first three bins (the best twins: 0-6th percentile, but also 0-12th, and 0-18th), each of which showed significantly better distance modulus scatter than the average pair of SNe. The inference (Section~\ref{sec:dicussion}) proceeds from the provided tables which are for the near-maximum analysis with the extinction $R_V$ fixed to 3.1. Table~\ref{tab:summarystats} shows summary statistics for each category of twin SNe.

\begin{deluxetable*}{cccc}[ht]
\tablehead{\colhead{Twins Percentile Range} & \colhead{Fraction of SNe with Any Twins} & \colhead{Observed \tdmean} & \colhead{Observed \tdmax}}
\startdata
0-6th & 38/49 = 0.78 & 59/1176 = 0.05 & 7/48 = 0.15 \\
0-12th & 43/49 = 0.88 & 59/588 = 0.10 & 1/4 = 0.25 \\
0-18th & 47/49 = 0.96 & 89/588 = 0.15 & 3/8 = 0.38  \\
\enddata
\caption{Summary statistics for the \citet{fakhouri15} data, showing the exact fractions, as well as decimal representation (rounded to two decimal places). \tdmean is the mean fraction of SNe each SN twins with: the mean number of twins divided by $\NSNe - 1$. \tdmax is the maximum twinning fraction: the maximum number of twins that any SN has divided by $\NSNe - 1$. \label{tab:summarystats}}
\end{deluxetable*}

\begin{figure*}[ht]
\begin{centering}
\includegraphics[width=0.51 \textwidth]{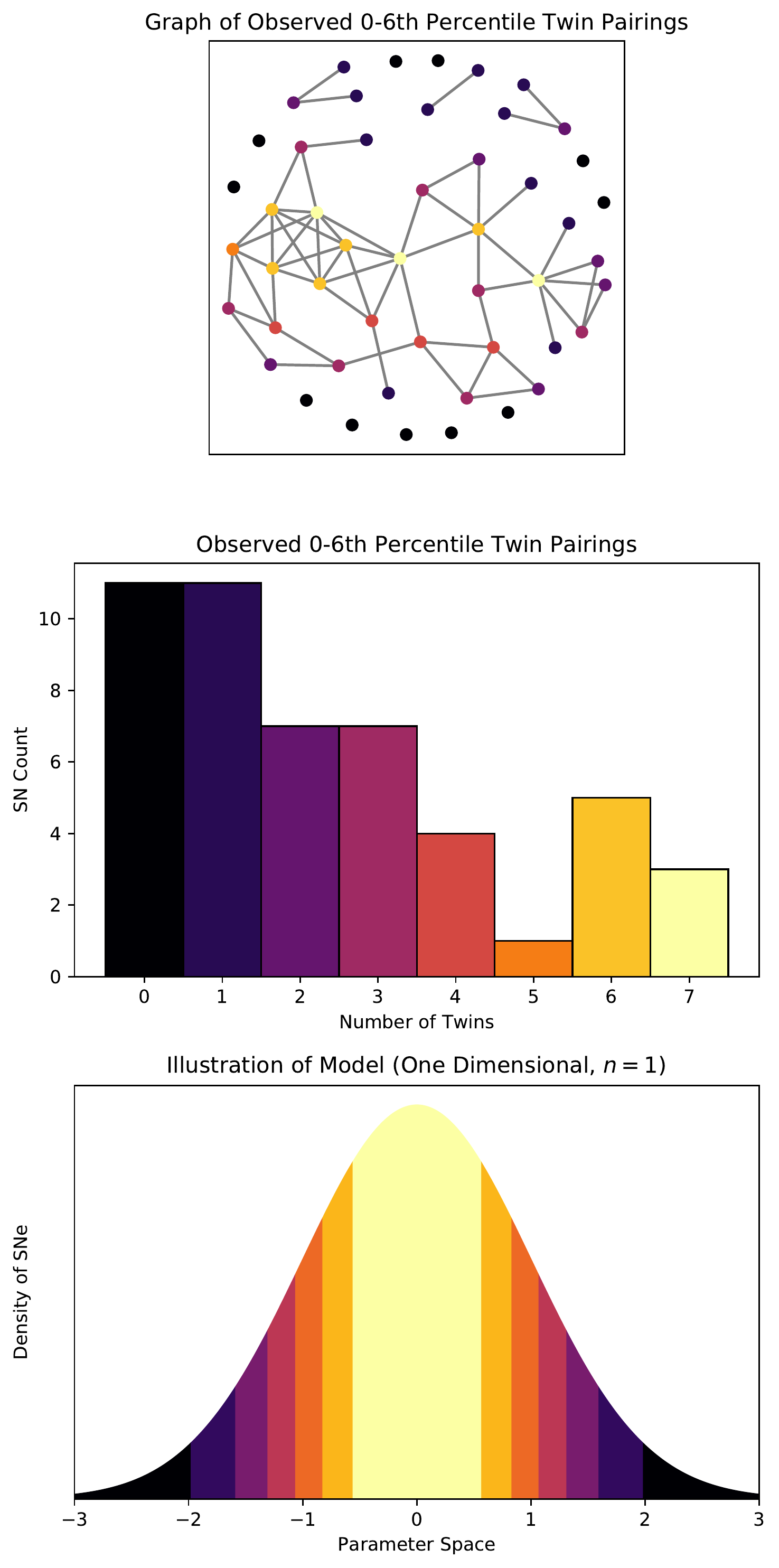}
\caption{Illustration of the data and the intuition behind the assumptions in this work. The {\bf top panel} shows a graph of 6th percentile twins. Each SN is color-coded by its number of twins; links indicate twins better than 6th percentile. The vertex locations are chosen to optimize a simple criterion that attempts to place twin SNe closer together than non-twin SNe, as \we hypothesize is true in parameter space. A clear gradient of twins density away from the center is visible, which \we take as evidence that a mode exists in parameter space. The {\bf middle panel} shows a histogram of the number of twins. Finally, the {\bf bottom panel} illustrates a Gaussian in parameter space; the mode of this distribution corresponds to the region with a high density of twins. This work constrains the dimensionality of the parameter space using the geometric property that higher dimensionality gives a more compressed range in parameter space (as the distribution narrows into a shell that is fractionally thin compared to its radius), and thus the distribution of the number of twins grows increasingly narrow.}\label{fig:twinsmesh}
\end{centering}
\end{figure*}

As noted above, this analysis probes the dimensionality of the SN Ia parameter space using the PDF of twin SN fraction (the PDF of the fraction of SNe each SN twins with). Figure~\ref{fig:twinsmesh} shows a qualitative overview illustrating the assumptions. The top panel presents a graph of the 0-6th percentile \citet{fakhouri15} twins pairings, in which twin SNe are plotted closer together on average than non-twin SNe. A mode of higher twins density is visible towards the center. The middle panel shows a histogram of the number of twins each SN has; the width of this distribution is broader than one would find if all SNe had the same expected number of twins. \We thus formulate the hypothesis illustrated in the bottom panel of  Figure~\ref{fig:twinsmesh}: there is a mode in parameter space, where there is an increased density of SNe (e.g., the core normals in the \citealt{branch06} classification). This mode thus corresponds to a region of SNe with higher numbers of twins, as the mean distance between SNe is smaller. Under this hypothesis, the dimensionality of the underlying parameter space can be measured using a general property of high-dimensional volumes: as the dimensionality increases, most of the volume concentrates near the surface. For example, for a hypersphere, the fraction of the volume in the inner 90\% is $0.9^n$, which rapidly approaches zero as the dimensionality $n$ approaches infinity. The same is true for Gaussian distributions; the sum of the squares of $n$ independent standard normal distributions (i.e., the radius squared) is distributed as $\chi^2(n)$, which has mean $n$ but standard deviation $\sqrt{2 n}$, so the relative width shrinks to zero as $n$ approaches infinity. Thus, as the dimensionality increases, the SNe increasingly come from regions of the parameter-space distribution away from both the mode and the periphery. Thus, the width of the predicted number-of-twins distribution shrinks as the dimensionality increases.

This approach is very complementary to nonlinear dimensionality discussed above. It naturally handles SNe with no twins by inferring such SNe are towards the periphery of the parameter distribution. However, this approach discards the network of twin pairs, focusing only on counting statistics. Thus it discards some information in exchange for a less rigid, probabilistic treatment of the data.

\begin{figure}[ht]
\begin{centering}
\includegraphics[width=0.5 \textwidth]{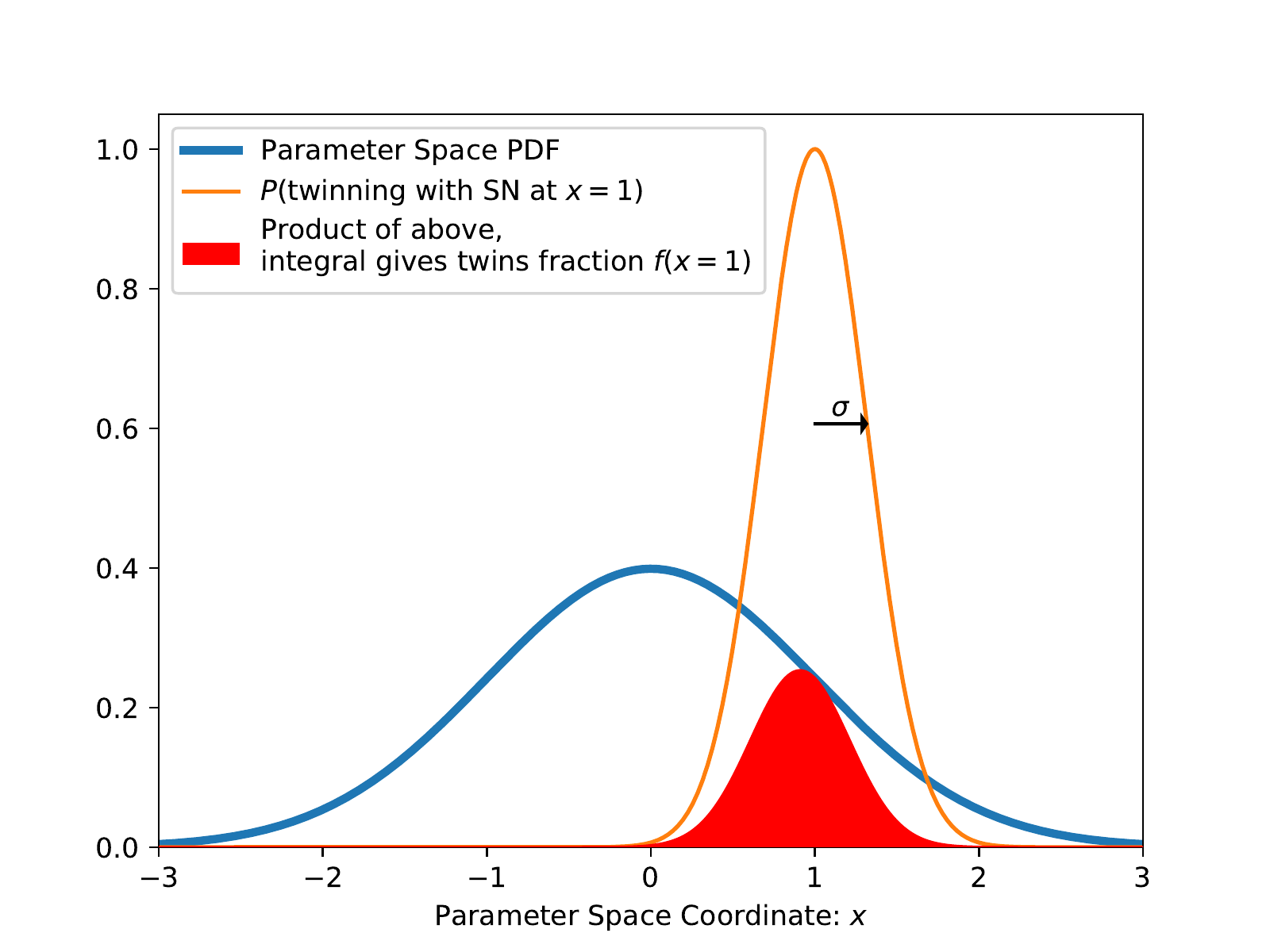}
\caption{Illustration of the model in 1D ($n=1$). The parameter space PDF is a unit normal, shown in blue. The probability that a SN twins with one at $x=1$ is shown in orange. This probability peaks at 1 if the two SNe have the same coordinates in parameter space (here, $x=1$), and falls off as an isotropic Gaussian of width $\sigma$. At a given point, the twins density (\td) is the integral over the product of these two functions, shown filled in red for a SN at $x=1$.}
\label{fig:illustration}
\end{centering}
\end{figure}

\subsection{Parameter Inference} \label{sec:inference}

\begin{figure}[ht]
\begin{centering}
\includegraphics[width=0.5 \textwidth]{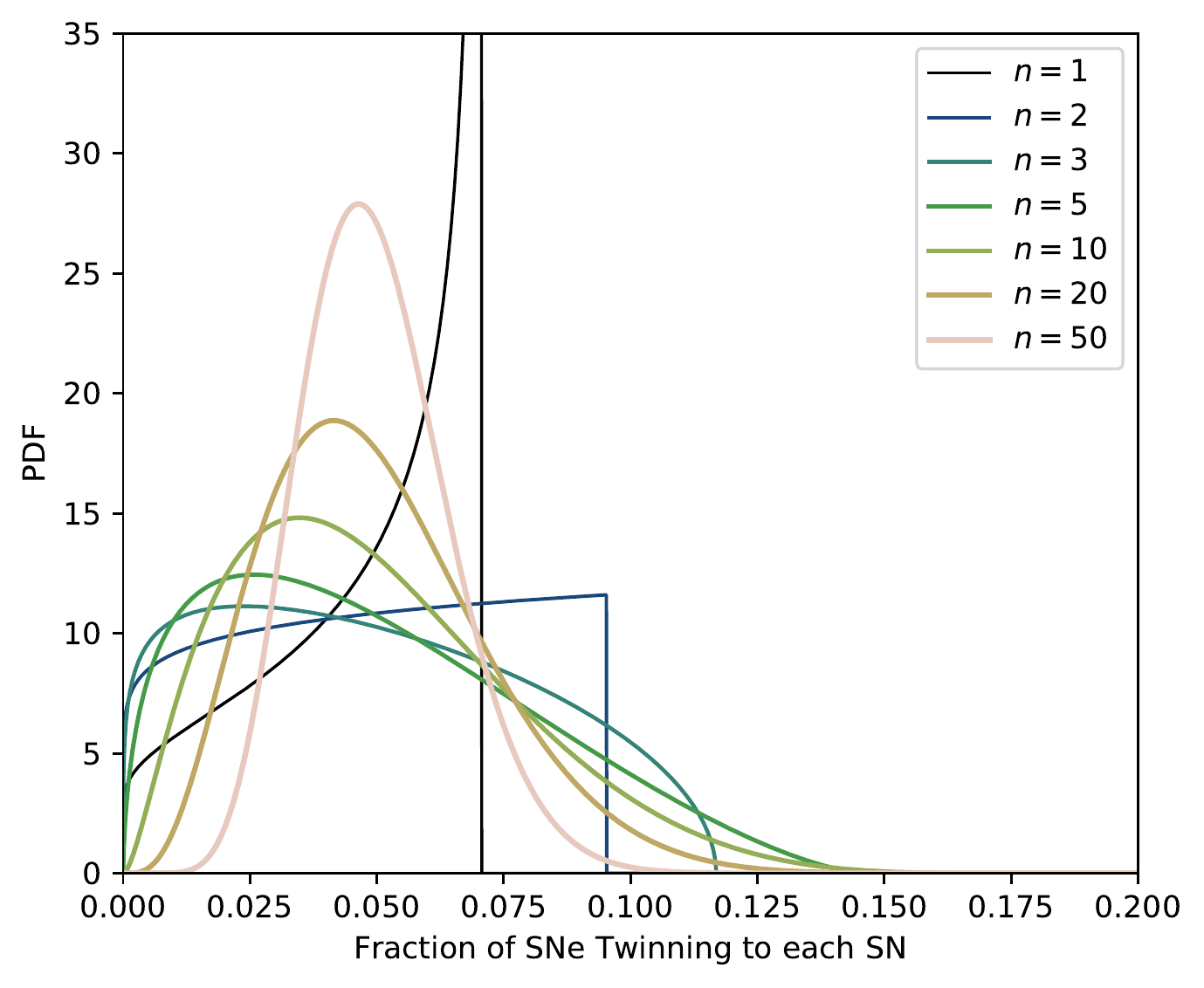}
\caption{PDF of fraction of SNe twinning as a function of the dimensionality (Equation~\ref{EQ:TWINPDF}). To isolate the effect of dimensionality ($n$), the shown PDFs fix the mean fraction to 0.05 (similar to that observed in the 0-6th percentile twins) by setting $\sigma$ according to Equation~\ref{eq:expectedfraction} for each value of \nd. The narrowing of the distribution for large \nd is apparent, as discussed in Section~\ref{sec:approach}.}
\label{fig:twinsPDFs}
\end{centering}
\end{figure}

The primary analysis assumes parameter space is an $n$-dimensional isotropic Gaussian, but Section~\ref{sec:dicussion} discusses extensions of this model. The other ingredient needed is the probability of two SNe twinning as a function of separation in parameter space. This probability should peak at 100\% for zero separation, then fall to zero as the separation gets large. \We assume this function is an $n$-dimensional isotropic Gaussian, with width $\sigma$. These assumptions are illustrated in Figure~\ref{fig:illustration}. These are strong assumptions, but they describe the data well (also discussed in Section~\ref{sec:dicussion}), and do it with only these two fit parameters ($n$ and $\sigma$). Interestingly, under these assumptions, the PDF of twin counts is analytic

\begin{align} 
\mathrm{PDF}(f \ |\ n, \sigma) = &\ \frac{ (1 + \sigma^2) \left[ (1 + \sigma^2) \log{\left[\frac{\tdmax}{\td} \right]} \right]^{\frac{1}{2}(n - 2) }}{   \left[ \frac{\tdmax}{\td} \right]^{\sigma^2} \tdmax\ \gammantwo  } \label{EQ:TWINPDF} \\
\tdmax \equiv &\ (1 + \sigma^{-2})^{-n/2}\;,
\end{align}
where \td is the fraction of SNe each SN twins with and \tdmax is the maximum twins density (realized for SNe at the center of the Gaussian parameter-space distribution). Appendix~\ref{sec:derivation} presents the derivation of Equation~\ref{EQ:TWINPDF}. Figure~\ref{fig:twinsPDFs} illustrates the behavior of the analytic PDF. For $n \le 2$, the distribution peaks at the maximum value of \tdmax. As discussed above, as the dimensionality gets very large, the width of the twins PDF decreases.

The uncertainties must also be considered before comparing against the \citet{fakhouri15} results. If the expected twins fraction (\td) is 2 out of 48 (i.e., $\NSNe - 1$) for a given SN, the observed number of SNe it actually twins with will be binomial-distributed as binomial(48, 2/48). Thus the inference must marginalize over the true (but not directly observed) expected number of SNe each SN twins with, transforming the inference problem from two parameters (\nd and $\sigma$) to $\NSNe + 2$ parameters:

\begin{align}
\td^{\mathrm{true}}_i &\sim \mathrm{PDF}(n,\ \sigma) \;,\ i \in 1,\ \ldots,\ \NSNe \label{eq:ModOne} \\ 
\td^{\mathrm{obs}}_i &\sim \mathrm{binomial}(\NSNe-1,\ \td^{\mathrm{true}}_i) \;,\ i \in 1,\ \ldots,\ \NSNe \;. \label{eq:ModTwo}
\end{align}
To perform this high-dimensional inference, \we use Stan \citep{carpenter17} through the PyStan  interface \citep{pystan}. Testing with simulated data (see below) indicated better frequentist coverage with a scale-invariant prior on $n$ (bounds are 1 to 50):
\begin{equation}
p(n) \propto 1/n \; \label{eq:prior}
\end{equation}
and a flat prior on $\sigma$ (bounds are 0.01 to 5).

\subsection{Simulated-Data Testing} \label{sec:simulated}

\begin{figure}[h]
\begin{centering}
\includegraphics[width=0.5 \textwidth]{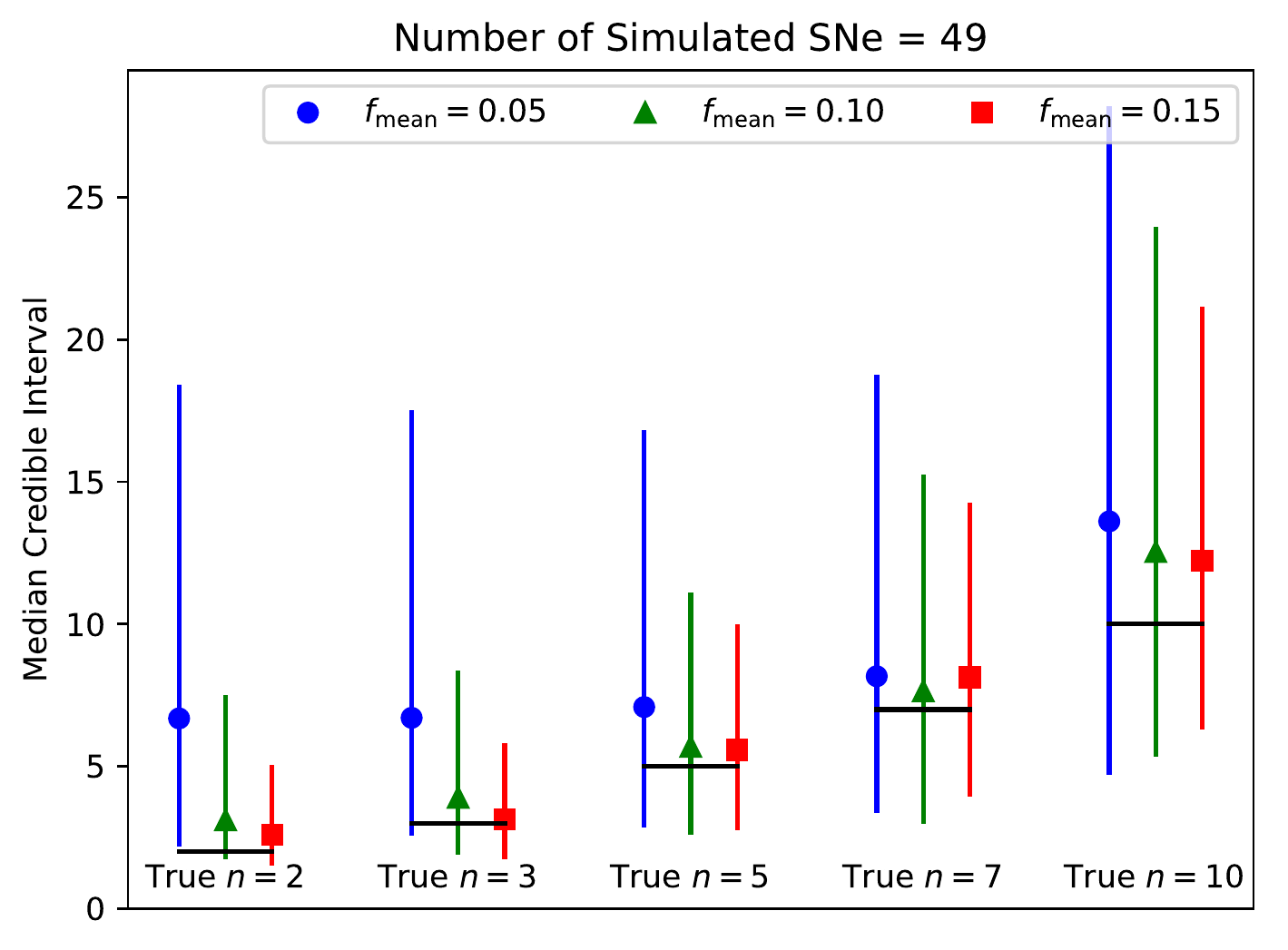}
\includegraphics[width=0.5 \textwidth]{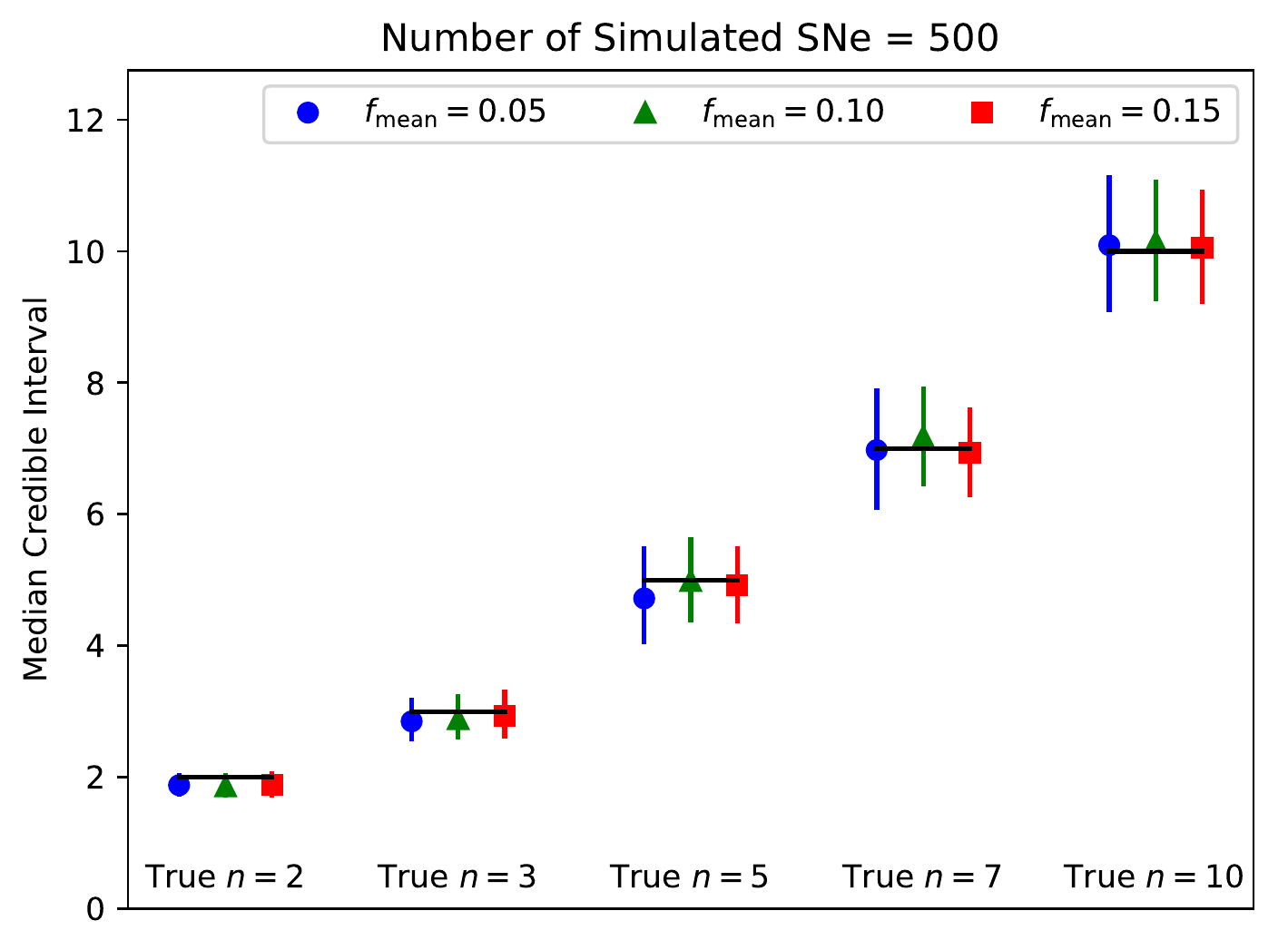}
\caption{Results of the simulated-data testing. The {\bf top panel} shows results for $\NSNe=49$; the {\bf bottom panel} shows results for $\NSNe=500$. \We run 100 simulated datasets for each of dimensionality ($n$) 2, 3, 5, 7, and 10, and \tdmean 0.05, 0.1, and 0.15. For each $n$/\tdmean pair, the lines show the median of the 68\% credible intervals from the posterior, and the points show the median of the posterior medians. The median credible intervals are grouped by $n$, with each \tdmean plotted with blue circles (0.05), green triangles (0.1), and red squares (0.15), respectively. The analysis shows generally mild biases.}
\label{fig:simdata}
\end{centering}
\end{figure}

This section describes the testing of the analysis on simulated data. \We simulate 3,000 datasets, or 100 for each combination of three parameters:
\begin{enumerate}
\item The dimensionality $n$, which has values 2, 3, 5, 7, and 10.
\item The average fraction of SNe each SN twins with (\tdmean), which has values 0.05, 0.10, and 0.15. This range corresponds roughly to the 0-6th, 0-12th, and 0-18th percentile twins in the actual data (Table~\ref{tab:summarystats}).
\item \NSNe can have values 49 (as with the real data) and 500 (to simulate the performance of this model on an expanded near-to-medium-term dataset).
\end{enumerate}
For each simulated dataset, \we infer the posterior on $n$. The median of the posterior serves as the best estimate, and the uncertainties are taken to be the difference between the percentile 84.1 and the median ($+ 1 \sigma$) and the median and percentile 15.9 ($-1\sigma$). Figure~\ref{fig:simdata} shows the median estimate and the median uncertainties. For $\tdmean=0.05$, the credible intervals are wide and bounded from below, so lower simulated values of $n$ do not strongly reduce the inferred $n$. However, for simulated datasets with $\tdmean=0.05$ and $n \geq 3$, most of the credible intervals do contain the input value of $n$. Simulated datasets with larger values of \tdmean show good recovery irrespective of $n$. For all simulated datasets with $\NSNe=500$, the analysis shows reasonable recovery of the input parameters. \We conclude that the inference is performing adequately.

Other than the choice of priors, one possibility for the mild biases seen in Figure~\ref{fig:simdata} is the assumed uncertainties. The simulated data show weak correlations of $-0.3$ to $+0.2$ between counts of different numbers; such correlations are caused by statistical fluctuations in the locations of SNe in parameter space. For example, if more SNe happen to be near the mode, this increases the number of SNe with high numbers of twins and decreases the number of SNe with low numbers of twins. \We ignore these correlations for the primary analysis; the simulations show reasonable recovery of the input parameters with 49 SNe and much better recovery as the number of SNe increases, as it will in the future. Section~\ref{sec:limitanalysis} discusses a simple model that does include correlations.

\section{Results and Discussion} \label{sec:dicussion}

\begin{figure}[h]
\begin{centering}
\includegraphics[width=0.5 \textwidth]{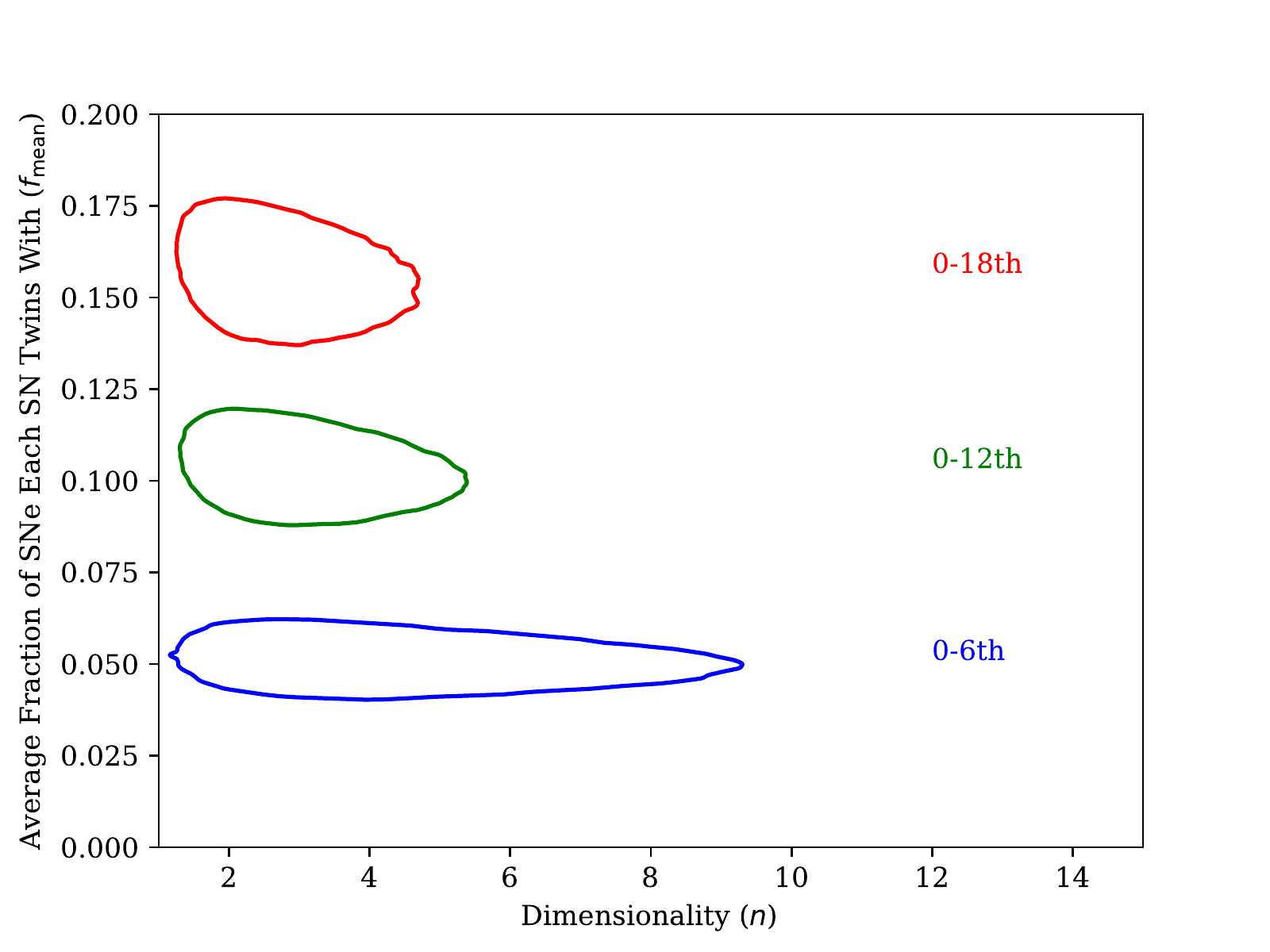}
\caption{68.3\% credible contours in \nd and \tdmean for the 0-6th percentile twins (bottom contour), the 0-12th percentile (middle contour) and 0-18th percentile (top contour). There is no evidence that more stringent twinness selections increase the dimensionality, although the analysis cannot rule out an increase given the size of the uncertainties.}
\label{fig:posteriorsixtwelveeighteen}
\end{centering}
\end{figure}

\begin{figure}[h]
\begin{centering}
\includegraphics[width=0.45 \textwidth]{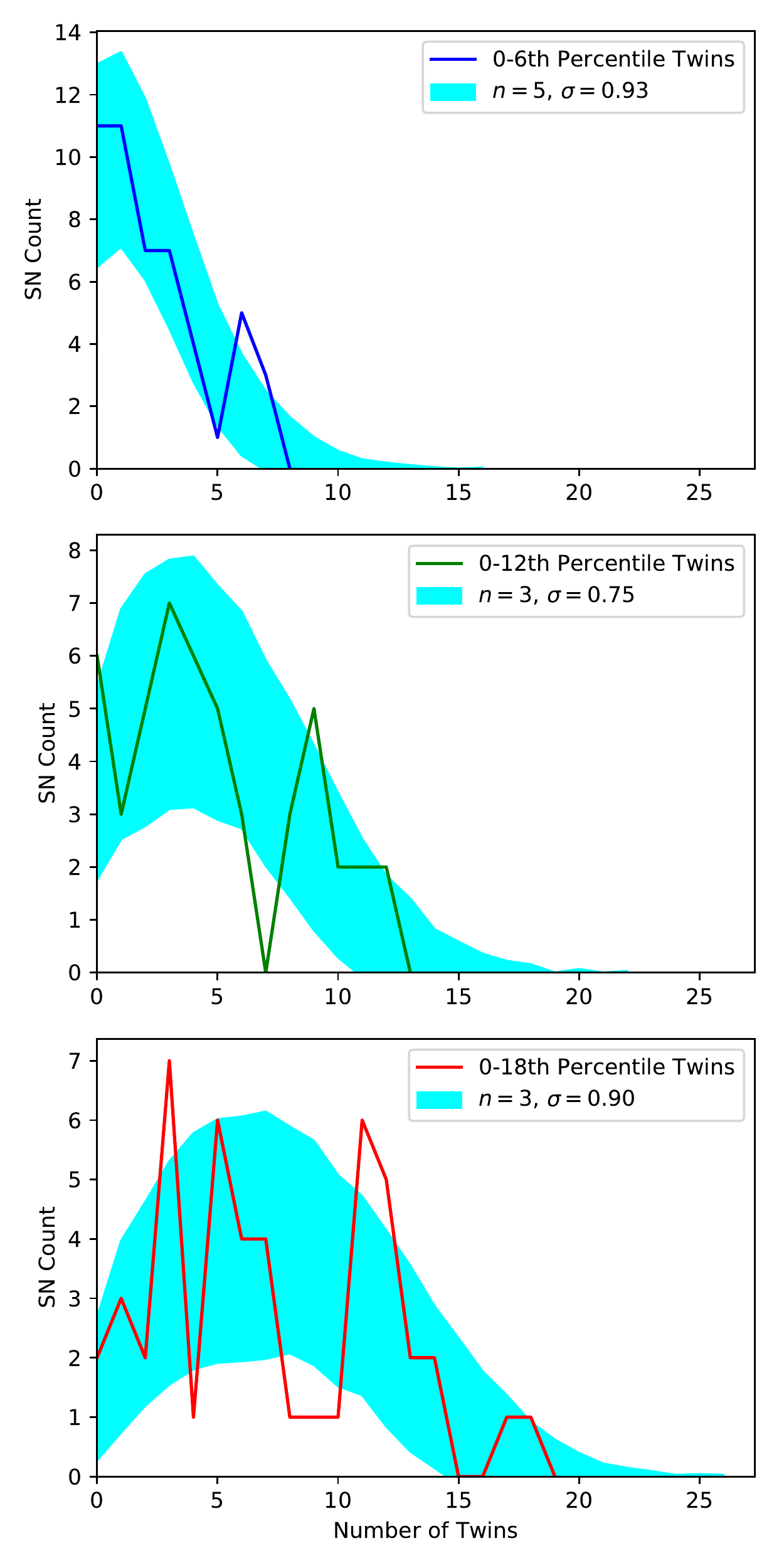}
\caption{The solid lines show the observed distribution of the number of twins for thresholds of 6th percentile ({\bf top panel}), 12th percentile ({\bf middle panel}), and 18th percentile ({\bf bottom panel}). As a comparison, the mean $\pm$ the standard deviation from 10,000 simulated datasets is overplotted on each (this comparison must be made to simulated data, rather than Equation~\ref{EQ:TWINPDF}, to add appropriate binomial noise). The model is an excellent match to the observed data.}
\label{fig:observedandsim}
\end{centering}
\end{figure}

\We now use Equations \ref{eq:ModOne}, \ref{eq:ModTwo}, and \ref{eq:prior} to infer $n$ and $\sigma$ for the \citet{fakhouri15} twin counts. Table~\ref{tab:credible} summarizes the parameter inference; in short, it shows a best estimate for $n$ of \bestestimatesofn, depending on the percentile range. Figure~\ref{fig:posteriorsixtwelveeighteen} shows the inferred credible contours, with one set of contours for each of the 0-6th, 0-12th, and 0-18th percentile twins.  Finally, Figure~\ref{fig:observedandsim} shows the best-fit model overplotted on the observed data. Instead of showing the analytic PDF (Equation~\ref{EQ:TWINPDF}), which would not be smeared by binomial noise, \we generate 10,000 simulated datasets for each percentile range and average them. Generating these datasets requires rounding the inferred $n$ to the nearest integer (\roundedn) and taking $\sigma$ conditional on those values.  In short, Figure~\ref{fig:posteriorsixtwelveeighteen} shows consistent $n$ estimates for the different twinness percentiles. while Figure~\ref{fig:observedandsim} shows that this model is an excellent description of the data. 

\begin{deluxetable*}{lcccc}[ht]
\tablehead{\colhead{Twins Percentile Range} & \colhead{$n$} & \colhead{$\sigma$} & \colhead{\tdmean} & \colhead{\tdmax}}
\startdata
0-6th & \nsixth & \sigsixth & \meanysixth & \ymaxsixth \\
0-12th & \ntwelveth & \sigtwelveth & \meanytwelveth & \ymaxtwelveth \\
0-18th & \neighteenth & \sigeighteenth & \meanyeighteenth & \ymaxeighteenth \\
\enddata
\caption{Inferred 68\% credible intervals, based on the 16th, 50th, and 84th percentile of the posterior samples. $n$ is the dimensionality, $\sigma$ describes the matching width, \tdmean is the mean fraction of SNe each SN twins with (a function of $n$ and $\sigma$, see Equation~\ref{eq:expectedfraction}), and \tdmax is the maximum fraction of SNe a SN can twin with (a function of $n$ and $\sigma$, see Equation~\ref{eq:SphereResult}). The inferred \tdmax values are lower than those directly seen in the data (Table~\ref{tab:summarystats}), as the binomial noise here is skewed positive, so the observed distribution is expected to exceed the model distribution. \label{tab:credible}}
\end{deluxetable*}

\subsection{Limitations}
There are some limitations to the inference as it currently stands. \We classify them into limitations of the data (Section~\ref{sec:limitdata}) and limitations of the analysis (Section~\ref{sec:limitanalysis}).

\subsubsection{Limitations of the Data} \label{sec:limitdata}

This analysis measures only the dimensionality of SNe that have passed the twins selection cuts of adequate phase coverage and smooth interpolation with a Gaussian process (as the SNe are intrinsically smooth in time, this cut rejects bad spectra). It is unlikely that an entire parameter would be completely excluded by these cuts (reducing the dimensionality by one). However, if the SN selection limited the range of one or more parameter, that would decrease the inferred dimensionality somewhat, as discussed in Section~\ref{sec:limitanalysis}.

The current inference is also sensitive only to parameters that affect spectrophotometry near maximum light, and only inside the rest-frame optical wavelength range considered in the \citet{fakhouri15} analysis (3300\ang to 8600\ang); see \citet{foley13} for two SNe which are similar in the optical, but different in the UV.

There is also no evidence either for or against a change in dimensionality with twinness percentile threshold. For example, it is possible that the top one percent of twins would give a different answer than the ranges considered here and in \citet{fakhouri15}; a much larger sample would be necessary to infer this.

\subsubsection{Limitations of the Analysis} \label{sec:limitanalysis}

One limitation of the current inference is the assumption of an isotropic distribution. For example, some dimensions may be much more important for twinning than others. To investigate the impact of this possibility, \we simulate an anisotropic distribution with a large number of dimensions, where each dimension has an exponentially decreasing RMS (RMS for dimension $i \propto e^{-i/\tau}$).\footnote{In practice, \we must simulate a number of dimensions $\gg \tau$. These results are based on simulating 30 dimensions and varying $\tau$ from 2 to 10.} For a range of $\tau$, the inference recovers $n \sim 1.4 \tau$, with a mild dependence on \tdmean. Thus, the results presented for the \citet{fakhouri15} data are consistent with a large number of parameters, as long as all but the first few have very little impact on twinness. Finally, \we note (although do not explore) that if the dimensionality of SNe Ia is heterogeneous, this analysis will infer an average, not the highest value. One way this situation may arise is if different progenitor channels (see \citealt{maoz14} for a review) have different dimensionality.

\begin{figure*}[ht]
\begin{centering}
\includegraphics[width=0.99 \textwidth]{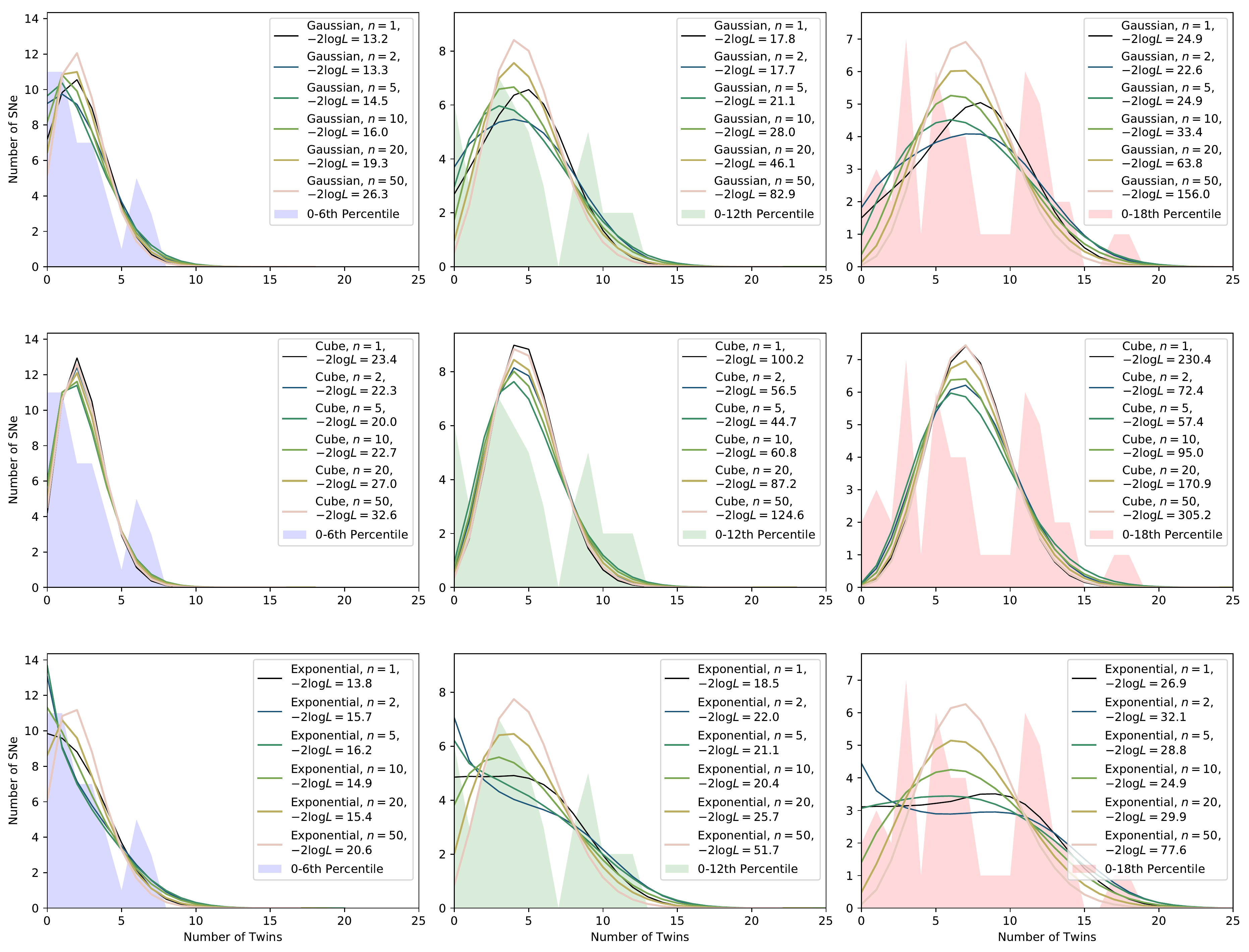}
\caption{Data simulations compared against the actual data. The rows show Gaussian ({\bf top}), hypercube ({\bf middle}), and exponential ($P(r) \propto e^{-r} \; r^{n-1} \; dr$,  {\bf bottom}), while the columns show 0-6th percentile twins ({\bf left}), 0-12th ({\bf middle}), and 0-18th ({\bf right}). The lines show increasing dimensionality ($n$) of the simulated data. For all distributions, the PDFs become narrower as the dimensionality increases, although the hypercube PDF is narrow even for low dimensionality. For each label, \we indicate an approximate $-2$ log likelihood ($\vec{r} \cdot C^{-1} \cdot r + \log{|C|}$); the log likelihoods are based on 100,000 simulations, so the differences between these values are determined to $\sim $ 1\%. The data simulations in the upper left are more similar to each other than the analytic PDFs (Figure~\ref{fig:twinsPDFs}), as these simulations have smearing due to binomial noise, as does the real data.\label{fig:otherPDFs}}
\end{centering}
\end{figure*}

Another limitation of the inference is the assumption of a Gaussian parameter distribution. Gaussians explain the data (including all three twinness thresholds) without any fine tuning of parameters. However, it is important to consider non-Gaussian parameter distributions. These distributions must be explored numerically, as no general analytic PDF exists equivalent to Equation~\ref{EQ:TWINPDF}. For each set of non-Gaussian data simulations, I compute an approximate $-2$ log likelihood
\begin{equation}
\vec{r} \cdot C^{-1} \cdot r + \log{|C|}\;, \label{eq:approxLL}
\end{equation}
where the residual $\vec{r}$ is the observed minus the mean of the simulations and $C$ is the covariance matrix of the simulations. Equation~\ref{eq:approxLL} is only approximate, as the distributions of twins counts are not quite Gaussian. A reasonable generalization of the $n$-dimensional Gaussian parameter distribution is the multidimensional S{\'e}rsic profile (\citealt{sersic63} presents the 1D version): $P(r) \propto e^{-r^{1/m}} \; r^{n-1} \; dr$, where the $r^{n-1}$ comes from the volume element in $n$-dimensional space. Treating $m$ as a nuisance parameter broadens the tails of the dimensionality distribution, although the inferred dimensionality does not increase. Interestingly, the best-fit $m$ is consistent with 1/2, i.e., a Gaussian distribution. Figure~\ref{fig:otherPDFs} compares three PDFs to the actual data: 1) the top row shows Gaussian distributions, 2) the middle row shows hypercube distributions, and 3) the bottom row shows radial exponential distributions ($P(r) \propto e^{-r} \; r^{n-1} \; dr$). The exponential distribution has both a sharper mode and heavier tails than a Gaussian, while the hypercube has a broad mode and no tails, so these are useful illustrations. The hypercube distribution is disfavored for any $n$, while exponential distributions do match the data for moderate $n$ ($\sim$ 5 to 10), although not as well as a Gaussian. Large values of $n$ are disfavored for both the exponential and Gaussian distributions, as both produce number-of-twins distributions that are too narrow. \We thus conclude that the choice of distribution does not strongly affect the best fit. This should be explored in future work, especially as the number of SNe in twins analyses increases.

\section{Conclusions and Future Work} \label{sec:conclusions}

This work presents a framework for inferring the dimensionality of the SN Ia parameter space based on the narrowing of the twin-SN PDF with increasing dimensionality. The dimensionality inferred is $\sim$ 3 or 5, depending on the twinness threshold chosen (summarized in Table~\ref{tab:credible}). \We validate the analysis on simulated data, and show that the results are reasonably stable under different assumptions.

\We now briefly revisit the list of applications in Section~\ref{sec:intro}.
\begin{enumerate}
\item \AppThree It is likely that SED models with only one linear parameter of variation (e.g., SALT2 and \textsc{SNEMO2}) underdescribe SNe Ia, missing important SED variation. On the other hand, \textsc{SNEMO15} (with fourteen parameters of variation, not including color) may be showing us the limits of linear dimensionality reduction, i.e., using fourteen linear components to capture a much smaller number of nonlinear parameters. \textsc{SNEMO7}, with six linear parameters of variation, may be the best current trade between accuracy and efficiency.

\item \AppTwo Without establishing exactly what the \bestestimatesofn SED parameters are, \we can only come to a tentative conclusion. It is plausible that two to four more parameters can be added to light-curve fitting, substantially reducing astrophysical systematics in current and near-term multi-band light curves (indeed, this is what SNEMO7 was designed to do, discussed further in \citealt{rose19}). See also \citet{kim13} and \citet{he18}, which present examples of light-curve fitting applied to larger numbers of principal components. However, \we caution that small effects may matter even if the majority of SED variation is captured. A small effect that correlates with age or metallicity (and thus, that may change with redshift) may still be important to measure.

\end{enumerate}

Future work \citep{boone19} will expand the twins analysis to more SNe, as well as improving the twinness criterion. In addition to increasing the number of SNe, obtaining more phase and wavelength coverage than the at-maximum data used here will be necessary for validating these results.

\acknowledgements

\We thank Kyle Boone and Greg Aldering for useful discussions. \We applied this method to observations taken with the University of Hawaii 2.2m telescope on Maunakea. The author wishes to recognize and acknowledge the very significant cultural role and reverence that the summit of Maunakea has always had within the indigenous Hawaiian community. We are most fortunate to have the opportunity to conduct observations from this mountain.

\software{
Matplotlib \citep{matplotlib}, 
Numpy \citep{numpy}, 
Pystan \citep{pystan},
SciPy \citep{scipy}, 
Stan \citep{carpenter17}
}

\clearpage

\appendix

\section{Derivation of Equation~\ref{EQ:TWINPDF} \label{sec:derivation}}

Without loss of generality, \we assume the isotropic \nd-dimensional Gaussian parameter distribution has width 1 in each parameter, and is centered on zero. Begin by writing this distribution in Cartesian coordinates:

\begin{equation} \label{eq:CartesianPDF}
\mathrm{PDF}(x_1, \ldots, x_n) = (2 \pi)^{-n/2} e^{-\frac{1}{2} [x_1^2 + \ldots + x_n^2]} \;.
\end{equation}

The expected fraction of twins ($\td$) at a given point $(x_1^{\prime}, ..., x_n^{\prime})$ is the integral over this PDF times the twins probability (c.f., Figure~\ref{fig:illustration}):

\begin{align} 
\td(x_1^{\prime}, ..., x_n^{\prime}) & =  \int \left[ (2 \pi)^{-n/2} e^{-\frac{1}{2} [x_1^2 + \ldots + x_n^2]} \right] 
\left[ e^{-\frac{1}{2 \sigma^2} [(x_1 - x_1^{\prime})^2 + \cdots + (x_n - x_n^{\prime})^2]} \right] dx_1 \, \cdots \, dx_n \label{eq:CartesianIntegral} \\
& =  (2 \pi \sigma^2)^{n/2} \int \left[ (2 \pi)^{-n/2} e^{-\frac{1}{2} [x_1^2 + \ldots + x_n^2]} \right] 
\left[ (2 \pi \sigma^2)^{-n/2} e^{-\frac{1}{2 \sigma^2} [(x_1 - x_1^{\prime})^2 + \cdots + (x_n - x_n^{\prime})^2]} \right] dx_1 \, \cdots \, dx_n \label{eq:CartesianProduct} \\
& = (2 \pi \sigma^2)^{n/2} \left[ (2 \pi (1 + \sigma^2))^{-n/2} e^{-\frac{1}{2 (1 + \sigma^2)} [x^{\prime 2}_1 + \ldots + x^{\prime 2}_n]} \right] \;,\label{eq:CartesianResult}
\end{align}
where Equation~\ref{eq:CartesianResult} follows because the integrals in Equation~\ref{eq:CartesianProduct} are the convolution of two $n$-dimensional Gaussian distributions. Transforming to spherical coordinates ($r$) and dropping the $^{\prime}$ gives:
\begin{equation} \label{eq:SphereResult}
\td(r) = \frac{ e^{-\frac{r^2}{2 (1 + \sigma^2)} } }{(1 + \sigma^{-2})^{n/2}} = \tdmax\; e^{-\frac{r^2}{2 (1 + \sigma^2)} }\;, 
\end{equation}

where \tdmax is the maximum possible twins density ($\tdmax \equiv (1 + \sigma^{-2})^{-n/2}$) which is realized in the center of the distribution.\footnote{\We pause briefly to note that, as expected in the limit of small $\sigma$, Equation~\ref{eq:SphereResult} scales as $\sigma^n e^{-\frac{r^2}{2}}$ or in other words, the volume of an \nd-dimensional region of size $\sigma$ times the PDF.}

The next step is compute the PDF of this twins density. \We find it conceptually easiest to compute the CDF. Inverting Equation~\ref{eq:SphereResult}, the fraction of SNe with twins density $< \td$ is given by the fraction of SNe at a radius $r$ greater than:

\begin{equation}
\mathrm{CDF}(\td\ |\ n, \sigma) \equiv P\left(r > \sqrt{2 (1 + \sigma^2)\log{\left[\frac{\tdmax}{\td} \right]}}\right) \;.
\end{equation}

The fraction of the SN PDF (Equation~\ref{eq:CartesianPDF}) that lies outside this $r$ is:
\begin{align}
\mathrm{CDF}(\td\ |\ n, \sigma) & = \int_{r = \sqrt{2 (1 + \sigma^2)\log{\left[\frac{\tdmax}{\td} \right]}}}^{\infty} \frac{e^{-\frac{1}{2} r^2}}{(2 \pi)^{n/2}} dr \; d\Omega \\
& = \int_{r = \sqrt{2 (1 + \sigma^2)\log{\left[\frac{\tdmax}{\td} \right]}}}^{\infty} e^{-\frac{1}{2} r^2} \frac{2^{1 - n/2}}{\Gamma(n/2)} r^{n-1} dr \\
& = \frac{\Gamma \left[ \frac{n}{2},\ (1 + \sigma^2) \log{(\tdmax/\td)} \right]}{\gammantwo }\;.
\end{align}

Taking the derivative of CDF$(\td\ |\ n, \sigma)$ with respect to \td gives the desired PDF (Equation~\ref{EQ:TWINPDF}).

\begin{equation}
\mathrm{PDF}(\td\ |\ n, \sigma) = \frac{ (1 + \sigma^2) \left[ (1 + \sigma^2) \log{\left[\frac{\tdmax}{\td} \right]} \right]^{\frac{1}{2}(n - 2) }}{   \left[ \frac{\tdmax}{\td} \right]^{\sigma^2} \tdmax\ \gammantwo  } \;.
\end{equation}

Finally, \we note that the expected fraction of SNe each SN twins with is given by:

\begin{equation} \label{eq:expectedfraction}
\tdmean = \int_{\td = 0}^{\td = \tdmax} \mathrm{PDF}(\td\ |\ n, \sigma) \; \td \; d \td = \left[\frac{\sigma}{\sqrt{2 + \sigma^2}}\right]^n \;.
\end{equation}

\end{document}